\documentclass[aps,prl,twocolumn,showpacs,preprintnumbers,superscriptaddress,dblfloatfix]{revtex4-1}
\usepackage{graphicx}   
\usepackage{dcolumn}    
\usepackage{bm}         
\usepackage{amssymb}    
\usepackage{setspace}
\usepackage{amsmath, amssymb, setspace}
\usepackage{array}
\usepackage{booktabs}
\usepackage{mathrsfs}
\usepackage{indentfirst}
\usepackage{slashed}
\usepackage{float}
\usepackage{lmodern}
\usepackage{hyperref}
\usepackage{xcolor}
\usepackage{multirow}
\usepackage{epstopdf}
\usepackage{ulem}
\usepackage{tabularx}
\usepackage[justification=raggedright]{caption}
\usepackage[flushleft]{threeparttable}
\usepackage{hyperref}
\usepackage{soul}

\begin{document}


\title{Transmuted Gravity Wave Signals from Primordial Black Holes}

\author{Volodymyr Takhistov}
\email[]{vtakhist@physics.ucla.edu}
\affiliation{Department of Physics and Astronomy, University of California, Los Angeles\\
Los Angeles, CA 90095-1547, USA}

\date{\today}

\begin{abstract}
Primordial black holes (PBHs) interacting with compact stars in binaries lead to a new class of gravity wave signatures that we explore.~A small $10^{-16} - 10^{-7} M_{\odot}$ PBH captured by a neutron star or a white dwarf will eventually consume the host.
The resulting black hole will have a mass of only $\sim0.5-2.5 M_{\odot}$, not expected from astrophysics. For a double neutron star  binary system this leads to a transmutation into a black hole--neutron star binary, with a gravity wave signal detectable by the LIGO-VIRGO network. For a neutron star--white dwarf system this leads to a black hole--white dwarf binary, with a gravity wave signal detectable by LISA. Other systems, such as cataclysmic variable binaries, can also undergo transmutations. We describe gravity wave signals of the transmuted systems, stressing the differences and similarities with the original binaries.~New correlating  astrophysical phenomena, such as a double kilonova, can further help to distinguish these events. This setup evades constraints on solar mass PBHs and still allows for PBHs to constitute all of the dark matter. A lack of signal in future searches could constrain PBH parameter space.\end{abstract}


\maketitle

{\bf Introduction --}
Primordial black holes (PBHs) can appear from early Universe dynamics and can account for all or part of the dark matter (DM)~\cite{Zeldovich:1967,Hawking:1971ei,Carr:1974nx,CHAPLINE:1975,GarciaBellido:1996qt,Khlopov:2008qy,Frampton:2010sw,Kawasaki:2012kn,Kawasaki:2016pql,Cotner:2016cvr,Carr:2016drx,Inomata:2016rbd,Inomata:2017okj,Georg:2017mqk}. Null results in experimental searches for conventional DM
particle candidates as well as the recent discovery of gravitational waves (GW) by the LIGO-VIRGO network \cite{Abbott:2016blz,Abbott:2016nmj,Abbott:2017vtc} have led to reinvigorated interest in PBHs, as it was pointed out \cite{Bird:2016dcv,Sasaki:2016jop,Clesse:2016vqa} that they could contribute to the observed GW signal (possible relevance of PBHs to GWs has been mentioned even earlier, e.g. \cite{Nakamura:1997sm,Clesse:2015wea,Bond:1984,Carr:1980}).

Primary signals expected \cite{Abadie:2010cf} in ground-based GW detectors like LIGO originate from neutron star -- neutron star (NS-NS), neutron star -- black hole (NS-BH) as well as black hole -- black hole (BH-BH) binary mergers.
Multiple BH-BH mergers have already been observed \cite{Abbott:2016blz,Abbott:2016nmj,Abbott:2017vtc}. Recently, a NS-NS GW signal GW170817 along with accompanying electromagnetic emission has also been detected \cite{TheLIGOScientific:2017qsa,GBM:2017lvd}.
The upcoming space-based LISA experiment \cite{LISA:2017} will allow for GW searches in the lower frequency domain, relevant, among others, for signals associated with white dwarf (WD) binaries.~The onset of GW astronomy has motivated a series of recent studies \cite{Raidal:2017mfl,Eroshenko:2016hmn,Sasaki:2016jop} on PBH merger signals as well.
Considered PBHs, assumed to be primary binary components (i.e. PBH -- PBH mergers), typically lie within the $\simeq 5-100 \, M_{\odot}$ mass range that is relevant for LIGO stellar-mass BH merger signals.~Solar mass PBHs \cite{Naselskii:1985} could also participate in mergers \cite{Nakamura:1997sm}.

The smallest known BHs reside in the $\sim5-10 M_{\odot}$ range \cite{shaposhnikov:2009} (see \cite{Kovetz:2016kpi} for mass distribution). In general, astrophysical BHs are not expected to have mass below the $2 \lesssim M_{\rm TOV}/M_{\odot} \lesssim 2.2$ boundary set by the  Tolman - Oppenheimer - Volkoff stability limit $M_{\rm TOV}$ for non-rotating neutron stars \cite{Rezzolla:2017aly,Ruiz:2017due}, with the lower cut-off coming from pulsar observations \cite{Antoniadis:2013pzd}. Uniform rotation allows for the maximum supported mass to reach $1.2 \, M_{\rm TOV}  (\simeq 2.6 M_{\odot})$~\cite{Breu:2016ufb}.
Above this limit the star collapses to a black hole (or hypothetically, a quark star \cite{Ivanenko:1965}). While for WDs the Chandrasekhar limit imposes a $\simeq 1.4 M_{\odot}$ stability bound that is lower, the expected result is a type Ia supernova explosion without a compact remnant \cite{Mazzali:2007et} (although, accretion-induced collapse results in a neutron star \cite{Fryer:1998jb}). It is possible, however, to naturally produce $\simeq 0.5-2.5 M_{\odot}$ BHs from compact stars that have captured a small PBH.

In this \textit{Letter} we explore how 
novel GW signatures from binaries with atypical solar mass BHs can act as probes of tiny PBHs constituting DM. If a $10^{-16}  \lesssim M_{\rm PBH}/M_{\odot} \lesssim 10^{-7} $ PBH interacts and is captured by a compact star (NS or WD), it will eventually consume the host \cite{Capela:2013yf}.~For neutron stars, this scenario implies that corresponding binaries will be ``transmuted" into binaries with a $\simeq 1.2-2.5 M_{\odot}$ BH instead of the NS, where the lower mass bound comes from observations \cite{Martinez:2015mya} and theoretically could be far smaller~\cite{Haensel:2002cia,Strobel:2000mg}.~Similarly, a ``transmuted'' white dwarf, which could be part of an interacting cataclysmic variable binary with a white dwarf or a main sequence (MS) star companion, will result in a $\simeq 0.5-1.4 M_{\odot}$ BH. What can the GW signals and accompanying counterparts from such systems tell us about PBHs?

{\bf Black hole capture --} The capture rate of PBHs on NS can be estimated following \cite{Capela:2013yf} and depends on the PBH mass $M_{\rm PBH}$, DM density $\rho_{\rm DM}$ and velocity dispersion $\overline{v}$ (assumed to follow Maxwellian distribution).~The full capture rate is given by $F = (\Omega_{\rm PBH} / \Omega_{\rm DM}) F_0$, where $\Omega_{\rm PBH}$ is the PBH contribution to the overall DM abundance $\Omega_{\rm DM}$.~Here, $F_0$ is the base NS capture rate in Milky Way Galaxy (MW) and is given by 
\begin{equation} \label{eq:f0calc}
F_0 = \sqrt{6 \pi} \dfrac{\rho_{\text{DM}}}{M_{\rm PBH}} \Big[\dfrac{R_{\rm NS} R_s}{\overline{v}(1 - R_s/R_{\rm NS})}\Big] \Big(1 - e^{- E_{\rm loss}/E_b}\Big)~,
\end{equation}
where $E_b = M_{\rm PBH} \overline{v}^2/3$, $R_{\rm NS}$ is radius of the NS with mass $M_{\rm NS}$ and Schwarzschild radius $R_{\rm S} = 2 G M_{\rm NS}$. Convention $c = 1$ is used throughout.~PBH will be captured if the interaction energy loss $E_{\rm loss} \simeq 58.8 \,G^2 M_{\rm PBH}^2 M_{\rm NS}/R_{\rm NS}^2$ exceeds its initial kinetic energy   \cite{Capela:2013yf}.~We assume that a typical NS is described by $R_{\rm NS} = 10$ km and $M_{\rm NS} = 1.5 M_{\odot}$ \cite{Shapiro:1983du}.~In NS-NS binaries the system mass, and hence the interaction rate, is twice that of a single NS. In the case of WD, the star mass $M_{\rm WD}\sim 1 M_{\odot}$ is lower, but the radius $R_{\rm WD}\sim10^3$ km is drastically larger \cite{Shapiro:1983du}. This leads to suppression of $E_{\rm loss}$ and the resulting WD capture rate is several orders below that of NS.~The total number of PBHs captured by a NS is given by $F t$, where $t$ is the relevant interaction time. 

Once gravitationally captured, PBH will settle inside the star and grow via Bondi spherical accretion \cite{Capela:2013yf, Kouvaris:2013kra,Markovic:1994bu,Shapiro:1983du}.~For typical NS, the time for captured PBH to settle within the star is \cite{Capela:2013yf} $t_{\rm set}^{\rm NS} \simeq  9.5 \times 10^3 (M_{\rm PBH}/10^{-11} M_{\odot})^{-3/2}$ yrs.~For WDs we find that
$t_{\rm set}^{\rm WD} \simeq  6.4\times10^6 (M_{\rm PBH}/10^{-11} M_{\odot})^{-3/2}$ yrs.~Once settled inside, the Bondi accretion time for BH to consume the star is given by $t_{\rm con} = v_s^3 / 4 \pi \lambda_s G^2 \rho_c M_{\rm PBH}$, where $v_s$ is the sound speed, $\rho_c$ is the central density and $\lambda_s = 0.707$ is the density profile parameter (for a star described by an $n = 3$ polytrope).~For NS with $\rho_c = 10^{15}$ g/cm$^3$ and $v_s = 0.17$ the consumption time is 
$t_{\rm con}^{\rm NS} \simeq 5.3 \times 10^{-3} (10^{-11} M_{\odot}/M_{\rm PBH})$ yrs. For WD with $\rho_c = 10^8$ g/cm$^{3}$ and $v_s = 0.03$ (consistent with \cite{Kouvaris:2010jy}) the consumption time is $t_{\rm con}^{\rm WD} \simeq 2.9 \times 10^2 (10^{-11} M_{\odot}/M_{\rm PBH})$ yrs.
Hence, a PBH of mass $\gtrsim10^{-11} M_{\odot}$ captured by NS or WD will settle and consume the star within $\sim 10^6$ yrs. Since the binary population synthesis models \cite{Voss:2003ep,OShaughnessy:2006uzj,Nelemans:2001hp} predict a typical merger time of $\sim10^6 - 10^{10}$ yrs, we can take the system to effectively contain a solar mass BH after the capture. Our conclusions are not significantly affected by the star's equation of state.

{\bf Binary transmutation --} 
Several binary systems can have a transmuted NS: NS-NS, NS-BH and NS-WD. Using [t$_\text{N}$] to denote the newly formed $\simeq 1.2-2.5 M_{\odot}$ BHs, the resulting binaries are: BH[t$_\text{N}$]-NS, BH[t$_\text{N}$]-BH and BH[t$_\text{N}$]-WD. In the transmuted BH[t$_\text{N}$]-BH system, the second (original) BH can in principle be of drastically larger size than BH[t$_\text{N}$]. In an unlikely scenario, NS-NS can undergo two subsequent transmutations to a BH[t$_\text{N}$]-BH[t$_\text{N}$] binary containing two solar mass BHs. Similarly, several systems can have a transmuted WD. We denote such newly formed $\simeq 0.5 - 1.4 M_{\odot}$ BHs with [t$_\text{W}$]. The WD-NS, WD-WD, WD-MS binaries will produce BH[t$_\text{W}$]-NS, BH[t$_\text{W}$]-WD as well as BH[t$_\text{W}$]-MS systems, respectively. In an improbable scenario WD-WD or NS-WD can undergo two subsequent transmutations, leading to a binary with two atypically small solar mass BHs.
For the rest of this work we shall focus primarily on transmutation of NSs and NS-NS systems, which have the best understood GW signal.

Previously~\cite{Fuller:2017uyd}, we have shown that up to $\sim0.1-0.5 M_{\odot}$ of neutron rich material could be ejected as a result of a PBH-NS interaction. Since the material escapes, the final transmuted black hole will be $\sim10-50\%$ less massive than the parent neutron star and comparable in mass to a WD. This deviation, and hence the amount of ejecta, can be tested with the resulting GW signal.

{\bf Merger time --} The binary systems are expected to merge due to gravitational wave emission within Galactic timescale (see \cite{Faber:2012rw, Shibata:2011jka,Postnov:2014tza} for review). For simplicity, we assume that the binary orbit is circular (eccentricity $e \simeq 0$).
The merger time $\tau_{\rm mgr}$ for two point masses $m_1$ and $m_2$ with an orbital period $P_{\rm orb}$ is given, to a good approximation, by   \cite{Peters:1963ux,Peters1964,Postnov:2014tza}
\begin{equation} \label{eq:tmerge}
\tau_{\rm mgr} \simeq 4.8 \times 10^{10} \Big(\dfrac{P_{\rm orb}}{\rm day}\Big)^{8/3} \Big(\dfrac{\mu}{M_{\odot}}\Big)^{-1}\Big(\dfrac{M}{M_{\odot}}\Big)^{-2/3}  \rm{yrs}~,
\end{equation}
where $M = m_1 + m_2$ and $\mu = m_1 m_2 / M$ is the reduced mass. Using Kepler's laws, the orbital period dependence can be exchanged for binary separation distance.~Assuming that the relevant characteristics of the original system, such as the orbit eccentricity and star separation distance \cite{Voss:2003ep,OShaughnessy:2006uzj,Nelemans:2001hp}, persist for the transmuted binary, the merger time  will stay intact.~However, if significant amount of material is ejected from the parent star during the interaction, the merger time of the resulting binary will be visibly altered.
For $0.5 M_{\odot}$ of material ejected from the NS, the merger time of the BH[t$_\text{N}$]-NS system will
be $\sim40\%$ larger than that of the original NS-NS.

{\bf Gravity wave signal --} The general features of the resulting merger GW signals can be stated as follows \cite{Faber:2012rw,Shibata:2011jka,Postnov:2014tza}.
The GW luminosity, which describes the energy loss due to gravitational radiation, is given by
\begin{equation}
L_{\rm GW} = - \dfrac{d E_{\rm GW}}{dt} =  \Big(\dfrac{32}{5}\Big) G^{7/3} (\mathcal{M}_c \pi f_{\rm GW} )^{10/3}~,
\end{equation}
where $f_{\rm GW} = 2/P_{\rm orb}$ is the frequency of the emitted GWs and $\mathcal{M}_c = \mu^{3/5} M^{2/5}$ is the ``chirp mass''. For a non-zero binary eccentricity, luminosity will also include contributions from higher GW harmonics luminosity \cite{Peters:1963ux}.
The time variation of the GW frequency is
\begin{equation}
\dot{f}_{\rm GW} = \Big(\dfrac{96}{5}\Big) G^{5/3} \pi^{8/3} \mathcal{M}_c^{5/3} f^{11/3}_{\rm GW}~.
\end{equation}
Averaging over the period and orientations, the ``characteristic'' GW strain amplitude is given by 
\begin{equation}
h = 1.5 \times 10^{-21} \Big(\dfrac{f_{\rm GW}}{ 10^{-3} {\rm Hz}}\Big)^{2/3}  \Big(\dfrac{\mathcal{M}_c}{M_{\odot}}\Big)^{5/3} \Big(\dfrac{D}{{\rm kpc}}\Big)^{-1}~,
\end{equation}
where $D$ is the distance to the source. 
Assuming that the relevant system quantities  do not change, we conclude that the general GW features of the transmuted binaries will resemble the original systems. As with merger time, significant mass ejection from PBH-star interaction will result in observable discrepancies. A $0.5 M_{\odot}$ mass ejection will produce a $\sim20\%$ smaller chirp mass, leading to a $\sim30\%$ smaller signal strain.

Following the inspiral phase, as the binary separation drastically decreases and becomes comparable to few star radii in size the objects descend onto one another in a merger phase. For a NS-BH system this is the time instabilities occur and the NS can be either tidally disrupted or fully swallowed by the BH before disruption can happen.~If NS is tidally disrupted an accretion disk can form around the BH. The presence of an accretion disk can be  estimated by comparing the tidal disruption (mass-shedding) onset distance $d_{\rm tid}$ with the inner-most stable circular orbit (ISCO) radius $r_{\rm ISCO}$. The tidal disruption distance can be approximated as \cite{Etienne:2007jg}
$d_{\rm tid} \simeq q^{-2/3} \mathcal{C}^{-1} G M_{\rm BH}$,
where $q = M_{\rm BH}/M_{\rm NS}$ is the binary mass ratio and $\mathcal{C} = G M_{\rm NS} / R_{\rm NS} \sim 0.2$ is the NS compaction, which depends on the star's equation of state. For a non-spinning BH the ISCO radius is given by $r_{\rm ISCO} = 6 G M_{\rm BH}$, which decreases to $r_{\rm ISCO} =  G M_{\rm BH}$ for a maximally-spinning Kerr BH.
Thus, for equal mass objects ($q = 1$) we can expect a possible accretion disk formation as the two quantities are comparable, as further confirmed by more careful studies (\cite{Taniguchi:2007aq}; see Fig. 15). A non-negligible spin of the resulting solar mass BH in the transmuted binary will have a strong effect on the disk formation details and the expected GW signature will be distinct \cite{Kyutoku:2011vz}. In a NS-NS system one typically expects that a disk of material will surround the final remnant (BH or massive NS) \cite{Faber:2012rw}.
The resulting disk from either BH-NS or NS-NS could act as a power source for short gamma-ray burst (sGRB) \cite{Faber:2012rw}. Systems with magnetized stars could also have an EM precursor signal \cite{Paschalidis:2013jsa}.

The NS-NS system will finally settle to a dynamically stable state in a post-merger (ringdown) phase \cite{Faber:2012rw}. Depending on the system factors, such as the rotational profile, the merger remnant could either promptly collapse to a BH or form either of the three star states: stable, supramassive, hypermassive. The supramassive state could be unstable to collapse, while the hypermassive state will eventually result in a delayed collapse to a BH. The latter provides a possible mechanism for a delayed sGRB, something not possible in a BH-NS system. In the case of BH-NS system \cite{Shibata:2009cn}, if NS is not tidally disrupted and swallowed whole by the BH the resulting waveform will have some qualitative resemblance of the BH-BH signal with a ringdown. On the other hand, if NS is tidally disrupted the GW signal will be quickly damped after the inspiral, suppressing the ringdown waveform.

We have highlighted some similarities and differences between the original and the transmuted merger signals, referring for further details to recent studies in NS-NS \cite{Shibata:2005ss,Sekiguchi:2011zd}, NS-BH \cite{Shibata:2007zm,Lackey:2011vz,Lackey:2013axa} as well as NS-WD \cite{Paschalidis:2011ez} simulations. 
While discriminating GW signal features such as tidal deformability exist \cite{Lackey:2011vz,Lackey:2013axa}, they mainly reside in the high GW frequency range above $\sim 500-1000$ Hz. Since aLIGO-VIRGO 
is more sensitive to the lower frequency inspiral signal, which is in general similar for $q \simeq 1$ binaries, identification of transmuted NS systems based purely on the resulting GW signal could be challenging. 

{\bf Signal detection --} GW signals from the NS-NS mergers are detectable in the $10-10^4$ Hz frequency domain of Advanced LIGO-VIRGO \cite{TheLIGOScientific:2014jea} (see also KAGRA experiment \cite{Aso:2013eba}). Large variation in predictions of associated merger rates, ranging $\sim0.4 - 10^3$ events/year \cite{Abadie:2010cf,Chruslinska:2017odi}, highlights significant uncertainties in binary population synthesis models. GW signals from the WD mergers are covered by the $10^{-4} - 10^{-1}$ Hz frequency band of LISA \cite{LISA:2017}. Around $\sim0.1-150$ NS-WD systems are predicted to be detectable above noise by LISA with one year of integration time \cite{Nelemans:2001hp, Cooray:2004ae}. GW signals from the corresponding transmuted binaries will also reside within these frequencies and we estimate below the transmuted signal sensitivity. 

The approximative number $N_G$ of accessible Milky Way
Equivalent Galaxies (MWEGs) within a horizon distance $D_{\rm h}$ is given by \cite{Abadie:2010cf}
\begin{equation}
N_G (D_{\rm h}) =  (1.16 \times 10^{-2}) \dfrac{4 \pi}{3} \Big(\dfrac{D_{\rm h}}{{\rm Mpc}}\Big)^3~,
\end{equation}
where the $1.16 \times 10^{-2}$ Mpc$^{-3}$ factor accounts for MWEGs density in space extrapolated from blue-light luminosity \cite{Kopparapu:2007ib}.~A more conservative count is obtained by inclusion of an additional $(1/2.26)^3$ correction factor that accounts for an all-sky optimization, which we exclude for the best reach estimate and include for the lower sensitivity bound calculation. The horizon distance defines
the maximum effective distance an optimally oriented source 
can be detected at the signal-to-noise ratio $S/N = \rho$ of 8. With Advanced LIGO-VIRGO network at design sensitivity, $D_{\rm h} = 445$ Mpc for NS-NS mergers \cite{Abadie:2010cf}. This $S/N$ choice is conservative and we set $\rho = 4$ for estimating the best reach, while still requiring strong signal discrimination. Since $\rho \sim 1/D$, this effectively rescales the reach distance to source $D$ by a factor of 2 and increases the total $N_G (D)$ by a factor of 8. Our choice keeps $D$ below the Gpc ($z \sim 0.2$) distance level, above which the effects \cite{Finn:1995ah,Abadie:2010cf} of cosmological redshift $z$, which we ignore, become important.

The total number of NS-NS systems transmuted in the Galaxy over time $t_t$ is $N_t = F t_t N_{\rm DNS}$, where $N_{\rm DNS}$ is the Galactic NS-NS binary population.~The Galactic merger rate can then be estimated \cite{Phinney:1991ei,Arzoumanian:1999,Lorimer:2008se} as $R_{\rm mgr} = N_t / \tau_p$, where $\tau_p = (\tau_s + \tau_{\rm mgr}) \lesssim 10$ Gyr is the total binary lifetime and $\tau_s$ is the spin-down time (i.e. ``characteristic age''). The number of expected transmuted events $N_{\rm exp}$ detected in observation time $t_{\rm obs}$, taking into account the detector's sensitivity reach via $N_G$, is $N_{\rm exp} = R_{\rm  mgr} N_G t_{\rm obs}$. 
The fraction of dark matter in the form of PBHs required to explain $N_{\rm obs}$ events observed in the experiment within $t_{\rm obs}$ is
\begin{equation}
\Big(\dfrac{\Omega_{\rm PBH}}{\Omega_{\rm DM}}\Big) = \dfrac{N_{\rm obs}}{(F_0 t_t N_{\rm DNS}/\tau_{p}) N_G t_{\rm obs}}~.
\end{equation}
To ensure that a binary is transmuted before the system merges we set $t_t = \tau_p$, canceling factors. For the detector running period we take one year. Further, we assume that even one observed transmuted event will be distinguished and take $N_{\rm obs} = 1$ to probe the PBH DM parameter space. The Galactic population of NS-NS binaries is fairly uncertain, with a predicted size of $N_{\rm DNS} = 7.5 \times 10^5$~\cite{Nelemans:2001hp} (see \cite{OShaughnessy:2006uzj} for a more conservative prediction).
\begin{figure}[tb]
\centering
\hspace{-2em}
\includegraphics[trim = 0.0mm .0mm 20mm 0mm, clip,width = 3.5in]{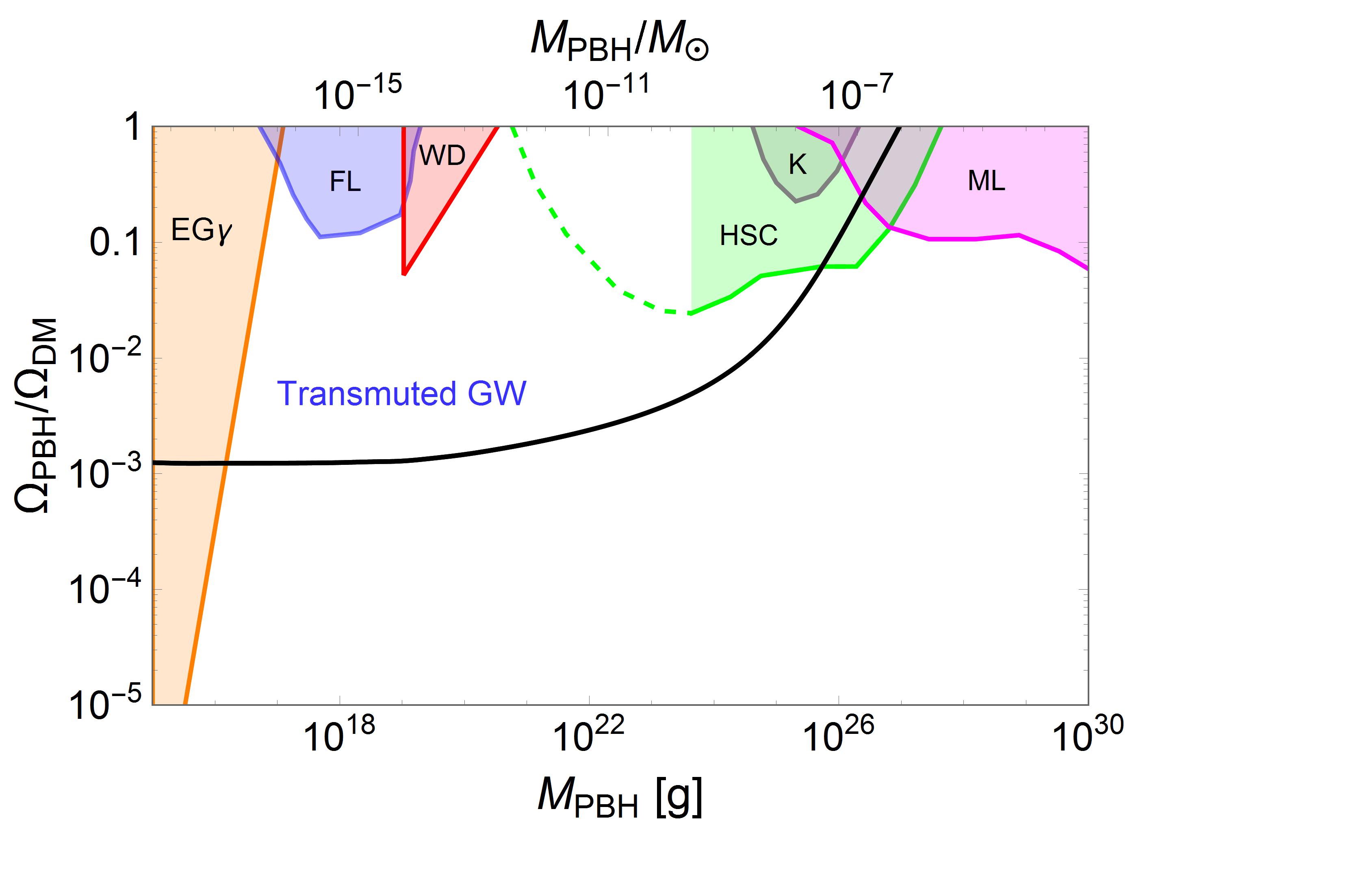}\\
\caption{
Parameter space where PBHs produce one or more transmuted GW signal events per year in aLIGO-VIRGO network running at design sensitivity. The black curve represents the signal rate calculated with the most optimistic input parameter choice ($N_G = 3.4 \times 10^7$, $N_{\rm DNS} = 7.5 \times 10^5$, $\rho_{\text{DM}} = 8.8 \times 10^{2}$ GeV/cm$^{3}$, $\overline{v}_{\rm NS} = 48$ km/s, $\overline{v}_{\rm DM} = 105$ km/s).
Constraints from extragalactic $\gamma$-rays from BH  evaporation~\cite{Carr:2009jm} (EG$\gamma$), femto-lensing~\cite{Barnacka:2012bm} (FL), white dwarf abundance~\cite{Graham:2015apa} (WD), Kepler star milli/micro- lensing~\cite{Griest:2013aaa} (K), Subaru HSC micro-lensing~\cite{Niikura:2017zjd} (HSC) and MACHO/EROS/ OGLE micro-lensing~\cite{Tisserand:2006zx} (ML) are displayed. Dashed line indicates that HSC
constraints are expected to be weaker than reported,
since PBH Schwarzschild radius becomes smaller than
the wavelength of light \cite{Takada:PBH2017}.}
\label{fig:PBHGW}
\end{figure}
For the best reach estimate we have assumed that 
binaries generally reside near the Galactic Center \cite{Pfahl:2003tf}, where the DM density is very high.
For the lower sensitivity bound calculation we have assumed that binaries are uniformly spread throughout the Galaxy. The other input quantities are varied as in \cite{Fuller:2017uyd}. 
The DM density spans $50~\text{GeV}/\text{cm}^{3} \leq \rho_{\text{DM}} \leq 8.8 \times 10^{2}$ GeV/cm$^{3}$.~The lower bound corresponds to the ``flat-core'' Burkert profile~\cite{Burkert:1995yz} with a uniform density in the central kpc region.~The upper bound is the volume-averaged maximum allowed mass of the DM within 0.1 kpc of the GC, which is set by the criterion that DM does not exceed the baryonic content of $\sim10^8 M_{\odot}$. 
We take DM velocity dispersion values to lie in the 
$50~\text{km/s} \leq \overline{v} \leq 200~\text{km/s}$ range. The lower $\overline{v}$ limit corresponds to a possible DM disk within the halo~\cite{Read:2008fh,Read:2009iv}, while the upper limit corresponds to the Navarro-Frenk-White DM density profile without adiabatic contraction~\cite{Kaplinghat:2013xca}.~We further have included the effects of natal pulsar kicks, modifying the capture rate $F_0$ as described in \cite{Fuller:2017uyd}. The considered pulsar velocity dispersion is between 48 km/s \cite{Cordes:1997my} and 80 km/s \cite{Lyne:1998}.
Our fit results are shown on Fig~\ref{fig:PBHGW}. The enclosed region above the black curve displays the parameter space for PBHs to constitute DM that can be probed with transmuted GW signals in the aLIGO-VIRGO network.
We note that the interactions of PBHs with masses below $\sim10^{-15} M_{\odot}$ are not well understood. 
For the most optimistic parameter choice the Galactic capture rate of a PBH by a single NS is $\sim 10^{-11}$/yr. Hence, with a typical NS-NS binary merger time of $\sim 10^6 - 10^{10}$ yrs transmuted events can contribute up to percent level of the total NS-NS GW event rate in any experiment. We stress that large astrophysical uncertainties can strongly affect the signal rate, with the most pessimistic choice for the  input parameters yielding a signal rate that is 9 orders lower than the result from the most optimistic parameter choice.

The Galactic NS-WD binary population is $2.2 \times 10^6$ \cite{Nelemans:2001hp}, which is higher than that of NS-NS. While the abundance of unresolved WD-WD binaries produces a stochastic ``confusion-background'' at frequencies below $\sim 10^{-3}$ Hz, a Galactic NS-WD merger as well as the corresponding transmuted binary GW signal are resolvable \cite{Nelemans:2001hp, Cooray:2004ae}.
Interacting cataclysmic variable WD-MS and WD-WD systems are also numerous in the Galaxy with a population of $4.2 \times 10^7$ \cite{Nelemans:2001hp}. Their signal, and hence the signal from the corresponding transmuted binaries with a WD-mass BH, can be discernible in LISA above the background \cite{Meliani:2000gz}. While we predict that the Galactic rates for such systems are very low, they provide unique signal possibilities.

We further comment on the signal rate from globular clusters as well as ultra-faint dwarf galaxies (UFDs). The population per unit stellar mass of binaries with a white-dwarf primary (cataclysmic variables or WD-WD) is enhanced by a factor of $\lesssim 10$ in globular clusters compared to the Galactic field due to stellar dynamics, with a combined population of $\mathcal{O}(10^{3-4})$ binaries in the full globular cluster system \cite{Benacquista:2011kv,Knigge:2011mq}. For binaries with a neutron star primary the full globular cluster system population can be estimated as $\mathcal{O}(10^{3})$ for NS-WD   and $\mathcal{O}(10^2)$ for NS-NS systems \cite{Benacquista:2011kv}. Observations of globular clusters show no evidence of sizable
dark matter content in such systems \cite{Bradford:2011aq,Ibata:2012eq}. In the UFDs the total stellar mass is significantly lower than that of the Galaxy, with the total star formation history of all 10 UFDs, as studied by \cite{Ji:2016} for r-process nucleosynthesis, amounting to only $\mathcal{O}(10^5) M_{\odot}$. Hence, the relevant number of binaries, which is highly uncertain, is not expected to be sizable (see e.g. \cite{Bramante:2016mzo}) relative to the Galactic population. The DM-dependent PBH capture rate in UFDs, already calculated in \cite{Fuller:2017uyd}, is similar (with a few order uncertainty) to the Galactic one. Thus, the Galactic transmuted GW signal rates, which depend on both the binary population as well as the DM content of the system, are not expected to be significantly altered by contributions from globular clusters or UFDs.

In our analysis we have assumed a monochromatic PBH mass function, allowing for a general study. While typically the PBH formation models predict an extended mass function, the details are highly model-dependent. A procedure for implementing an extended mass function, corresponding to a particular formation model, is outlined in \cite{Carr:2017jsz}.

{\bf Astrophysical signatures --}
Accompanying astrophysical signatures, such as electromagnetic transients, will provide additional handles for discerning signals associated with transmuted binaries.~Kilonovae \cite{Kasen:2014toa,Hotokezaka:2015eja,Li:1998bw,Metzger:2010,Roberts:2011,Piran:2012wd,Barnes:2013wka,Martin:2015hxa}, fast radio bursts (FRBs) \cite{Fuller:2014rza,Lorimer:2007qn,Katz:2016dti} and positrons not associated with a merger GW emission are characteristic of a star transmuted by a PBH \cite{Fuller:2017uyd}. As the transmuted binary later merges, a second set of astrophysical signatures, now associated with a binary merger GW signal, will be present. Like the original NS-NS, a transmuted BH[t$_{\rm N}$]-NS binary is also expected to result in a kilonova (\cite{Kasen:2014toa}; Fig. 9). Hence, if two kilonovae could be associated together, one without an accompanying merger GW signal and a delayed one with a GW signal, this will provide evidence for a transmuted binary system at play. 
Initial follow-up searches for electromagnetic transients associated with GW sources have already taken off \cite{Doctor:2016gdi} and near-future experiments, like the LSST \cite{Abate:2012za}, will allow to identify such coincidence signals in high detail \cite{Rosswog:2016dhy}. For coincidence signals associated with WD binaries see \cite{Thompson:2009}.

{\bf Stochastic background --}
Many sources and early Universe phenomena, including unresolved binary coalescences \cite{Rosado:2011kv}, can contribute to the stochastic GW background \cite{Allen:1996vm,Binetruy:2012ze} that is already being probed by LIGO/VIRGO \cite{TheLIGOScientific:2016dpb}. Stochastic GW background from PBHs has also been considered \cite{Clesse:2016ajp}.
An estimate for the background contribution from transmuted binaries can be made using the results of \cite{Phinney:2001di}. The present-day energy density per logarithmic GW frequency  interval divided by the critical density $\rho_c$ is given by 
\begin{equation}
\Omega_{\rm gw} = 1.3 \times 10^{-17} \Big(\dfrac{\mathcal{M}_c}{M_{\odot}}\Big)^{5/3} \Big(\dfrac{f_{\rm GW}}{10^{-3}{\rm Hz}}\Big)^{2/3} \Big(\dfrac{N_0}{{\rm Mpc}^{-3}}\Big) z_{\rm rs}~,
\end{equation}
where $N_0$ is the present-day comoving number density of merged remnants and $z_{\rm rs} = \langle (1+z)^{-1/3}\rangle/0.74$ accounts for the redshift contribution, evaluated for flat Universe, and is not very sensitive to the details of star formation history.~Transmuted binaries alter the already present binary population and their respective contribution to the background.~Compared to regular binaries the transmuted system count is small and since the mass and the general GW characteristics bare similarities to the original systems we do not expect significant changes to the already present binary GW background.

{\bf Conclusion --}
In this work we have shown how unusually small $\sim0.5-2.5 M_{\odot}$ BHs can be naturally produced in binaries and act as probes of minuscule $10^{-16} - 10^{-7} M_{\odot}$ PBHs contributing to DM, which are difficult to study otherwise. We explored a new class of GW signals associated with systems where a NS or a WD has been transmuted into a solar mass BH, commenting on similarities and differences with respect to the original (un-transmuted) binaries. Specific system characteristics, such as the amount of ejected mass from the star's transmutation, will have a visible effect on the resulting GW signal. Upcoming experiments, such as aLIGO-VIRGO and LISA, can detect these events and potentially probe a portion of parameter space for PBHs to constitute DM. Observation of correlating astrophysical phenomena, such as a double kilonova, will further help identifying them.
Transmuted systems are not expected to significantly alter the already existing stochastic GW background from the binaries. Finally, solar mass BH binaries from our setup evade the constraints on solar mass PBHs and thus still allow for PBHs to constitute all of DM.

Our work sets up the problem and calls for future
simulations of star-PBH interactions and associated binary mergers, which will further shed light on the details of the unique signals that we have outlined.

~\newline
NOTE:~During completion of this work, a related study of \cite{Bramante:2017ulk} has appeared. The authors, primarily focusing on NS-implosions from particle DM, looked at some of the same questions as we did.
Our study provides the first analysis and details of the associated GW production and signal, closing the gap and complementing the above work. Further, we have identified that WD capture as well as ejected mass will result in binaries with an even smaller solar mass BHs. Contrary to their work, we were able to find a portion of PBH DM parameter space that can be tested with the upcoming GW experiments.

{\bf Acknowledgments --}
We thank Alexander Kusenko for fruitful discussions and comments. This work was supported by the U.S. Department of Energy Grant No. DE-SC0009937. \(\)

\bibliography{BHNSGW}
\end{document}